# Phase diagram of the resistive state of the narrow superconducting channel in the voltage-driven regime

Y.S. Yerin, V.N. Fenchenko

*B.Verkin Institute for Low Temperature Physics and Engineering
of the National Academy of Sciences of Ukraine, 47 Lenin Ave., Kharkov 61103, Ukraine*
E-mail: yerin@ilt.kharkov.ua

E. Il'ichev

*Institut für Photonische Technologien, Albert-Einstein-Straße 9, Jena, 07745, Germany*

## Abstract

Based on the numerical solution non-stationary Ginzburg-Landau equations, we investigated the evolution of the order parameter of superconducting channels with different lengths under applied voltage (so-called voltage-driven regime). We calculated current-voltage characteristics for channels of different lengths and found out the origin of theirs characteristic disorder oscillations. For very long channels in the certain voltage interval we reveal the chaotic dynamic of the order parameter. Collected data allowed us to plot the most complete and detailed phase diagram of the resistive state of the superconducting channel in the voltage-driven regime.

## 1. Introduction

The Ginzburg-Landau theory predicts that below the critical temperature $T_c$ a narrow channel, depending on the value of the current flowing through it, is either in a homogeneous superconducting phase or in the normal state. By the narrow channel we assume such a quasi-one-dimensional system, the transverse dimensions of which are less than or on the order of the coherence length and the depth of penetration of the magnetic field. In this case, we also assume that the length of the channel is sufficiently large and exceeds the depth of penetration of the electric field into the system.

However, the experiments showed that above the Ginzburg-Landau critical current the superconducting state in the quasi-one-dimensional channel does not disappear completely, but is transformed into the so-called resistive state, in which the superconducting and normal regions coexist simultaneously [1,2]. Despite the fact that the superconductivity in the system did not completely collapse, the presence of normal domains leads to the appearance at the ends of the channel of a non-zero potential difference. Moreover, in this case we also observe electromagnetic radiation with characteristics similar to Josephson ones [2, 3].

The latter circumstance is direct evidence that the resistive state in narrow superconducting wires has non-equilibrium nature. Therefore, for accurate description it is necessary to use the appropriate kinetic equations. However, due to the complexity and bulkiness of the mathematical approach that accompanies their analysis, dynamic problems of superconductivity are often solved using the simpler non-stationary (time-dependent) Ginzburg-Landau (GL) equations, which are derived from the above kinetic equations in the limit of the critical temperature of the superconductor.

An important advantage of the time-dependent Ginzburg-Landau equations is their visual clarity, since for the description of non-equilibrium superconductivity order parameter is used, which reflects the macroscopic quantum properties of the superconductor.

For this reason the problem of stability of current states in a superconductor in the resistive phase has been investigated using time-dependent Ginzburg-Landau equations since the 70s of the last century. In the framework of phenomenological formalism it has been shown that to preserve the macroscopic coherence it needs in certain parts of the system that the order parameter phase have to slip by $2\pi$ with a simultaneous vanishing of its modulus [4]. Such areas, comparable in size to the coherence length, were called phase slip centers (PSC), in the case of a quasi-one-dimensional superconductor, and phase slip lines (PSL), if the superconducting system is a wide thin film (a two-dimensional object).

Traditionally, the existence of such topological defects in a superconducting system, whether it is a quasi-one-dimensional channel or a two-dimensional wide thin film, is detected by observing the jumps and their temperature dependence on the current-voltage characteristics (CVC) of the samples [2, 5]. With the appearance of low-temperature laser scanning microscopy it became possible to visualize the dynamics of the resistive state of a superconductor, namely, to observe the appearance of normal domains in it and keep track of how the pattern of the state changes with changing external conditions (temperature, voltage, current). Relatively recently, similar studies of the evolution of the resistive state and, therefore, the dynamics of PSL were carried out for a wide thin film of tin in the current-driven regime [6].

Numerical analysis of the GL equations shows that the character of formation of PSC in a quasi-one-dimensional system depends on the conditions that create the resistive state of the superconductor (the voltage-driven regime or the current-driven regime), and on the length of the superconducting channel. Traditionally, the vast majority of publications, both theoretical and experimental, are devoted to the investigation of the dynamics of PSC produced in a current-driven regime. However, as shown in a number of theoretical studies, in the voltage-driven regime certain features may appear in the dynamic behavior of the order parameter of the quasi one-dimensional superconductor, which are not present in the current-driven regime, but are reflected in the CVC of such systems.

In particular, the latter experiments have shown [7] that the superconducting narrow channel in the voltage-driven regime has a non-trivial S-shaped feature. In the same paper, using the time-dependent GL equations with a non-zero gap, the authors provide a theoretical background of this remarkable behavior of the CVC. Moreover, based on of numerical solution of the equations for a depairing factor much smaller than one, disordered current density fluctuations superimposing on the CVC of the channel were discovered.

As for the dynamic behavior of the order parameter in the voltage-driven regime, its theoretical analysis has been carried out in Ref. 8. There, using the numerical solution of the time-dependent gapless (the depairing factor equals infinity) GL equations, the authors tracked the nature of formation of PSC in the narrow superconducting channel without investigations the dependence of the transition to different modes of behavior of the PSC on the length of the channel and the voltage range.

The authors of Ref. 9 were able to refine the quantitative dynamics of the resistive state of the superconducting wire in the voltage-driven regime. In the framework of the time-dependent GL equations, they built a phase diagram indicating for which lengths and voltages we should expect the transition from the single PSC regime to the 2 PSC regime, and from the 2 PSC regime to the region where the solutions of the GL equations become more complex. However, they were not able to describe the dynamics of the order parameter in this region.

In this paper, based on the numerical solution of the time-dependent GL equations, we investigate the dynamics of the resistive state of quasi-one-dimensional filaments of different lengths in the voltage-driven regime with a visualization of the behavior of the order parameter modulus. Based on these data, a detailed version of the phase diagram of such a system has been built as a function of the length and the value of the voltage applied to it, compared with analogous characteristics given in Ref. 9. In this diagram we mark not only those voltage–length intervals, for which one or two PSC form in the channels, but we also indicate the area where the disordered nature of PSC formation with chaotic time-dependent behavior of the order parameter is observed. In addition, we propose a qualitative theory that explains the appearance of current density oscillations in the CVC of quasi-one-dimensional narrow superconducting channels.

## 2. Basic equations

The time-dependent GL equations for the investigated system in dimensionless units have the form:

$$u(\partial_t \psi + i\Phi \psi) - \partial_x^2 \psi - \tau \psi + |\psi|^2 \psi = 0 \tag{1}$$

$$j = -\partial_x \Phi - i(\psi^* \partial_x \psi - \psi \partial_x \psi^*). \tag{2}$$

In the above equations, the length is measured in units of coherence length $\xi_0 = \sqrt{\dfrac{\pi \hbar D}{8 k_B T_c}}$ for $T = 0$, where $D = \dfrac{1}{3} v_F l$ is the diffusion coefficient, and the time is in units of $t_0 = \dfrac{\pi \hbar}{8 k_B T_c}$. Moreover, $\psi$ is a complex value, a dimensionless order parameter normalized to $\psi_0 = \sqrt{\dfrac{8\pi^2 k_B^2 T_c^2}{7\zeta(3)}}$, $\Phi$ corresponds to the electrostatic potential, which is measured in units of $\dfrac{\hbar}{2 e t_0}$, $j$ is the density of current flowing through the system is composed of additive contributions of the normal quasiparticle flow and the Cooper condensate and is expressed through $j_0 = \dfrac{c\Phi_0}{16\pi^2 \lambda_0^2 \xi_0}$, where $\lambda_0$ is the London penetration depth of the magnetic field at $T = 0$. The temperature is introduced in the equations $\tau = 1 - \dfrac{T}{T_c}$, which will be assumed to be $0.1$ $(T = 0.9 T_c)$.

Parameter $u$ is a numerical parameter depending on the superconducting properties of the material, which is equal to the ratio of the relaxation time of the order parameter modulus to the relaxation time of its phase. According to the microscopic theory, $u$ takes on values depending on the degree of impurities of the superconductor. If $t_s T_c \ll 1$, where $t_s$ is the relaxation time by magnetic impurities, then $u = 12$. If $t_{imp} T_c \ll 1$, where $t_{imp}$ is the time of scattering by the impurities, the parameter $u$ is set to equal $\dfrac{\pi^4}{14\zeta(3)} \approx 5.79$. However, strictly speaking, the assumption of a wide range of values of $u$ does not contradict the microscopic theory, so there are no restrictions on the choice of an arbitrary positive value for this parameter. For this reason we have chosen $u = 1$.

Equations (1) and (2) must be supplemented by boundary and initial conditions corresponding to the voltage-driven regime. Based on the selected set of normalizing parameters, these conditions are as follows:

$$\psi(0,t) = |\psi^{(0)}|, \quad \psi(L,t) = |\psi^{(0)}| \exp(iVt), \tag{3}$$

$$\Phi(0,t) = 0, \quad \Phi(L,t) = V, \tag{4}$$

$$\psi(x,0) = |\psi^{(0)}|, \quad \Phi(x,0) = V\frac{x}{L}. \qquad (5)$$

where $L$ is the length of the channel, $|\psi^{(0)}| = \sqrt{\tau}$ is the equilibrium value of the order parameter modulus, and $V$ is the voltage applied.

The time-dependent Ginzburg-Landau equations (1) and (2) with the boundary and initial conditions (3)–(5) are solved numerically by the fourth-order Runge-Kutta method with the substitution of the time and space derivatives with finite-difference schemes. In the process of numerical simulation a time step was chosen to equal 0.01, and the minimum size of the spatial grid was 0.5, i.e., half of the coherence length of the first order parameter for $T=0$. Also note that to obtain CVC, current density averaging occurred in the time interval between 500 and 10 000. This initial value was chosen to avoid any influence of various kinds of relaxation processes occurring in the channel initially.

### 3. Results and discussions

A characteristic feature of a superconducting channel in the voltage-driven regime is the S-like shape of the CVC. In the process of numerical solution of Eqs. (1)-(5) we found that this feature does not hold for all systems, but only for those whose length exceeds the specified value of $L^{(S)} \approx 21$ (fig. 1).

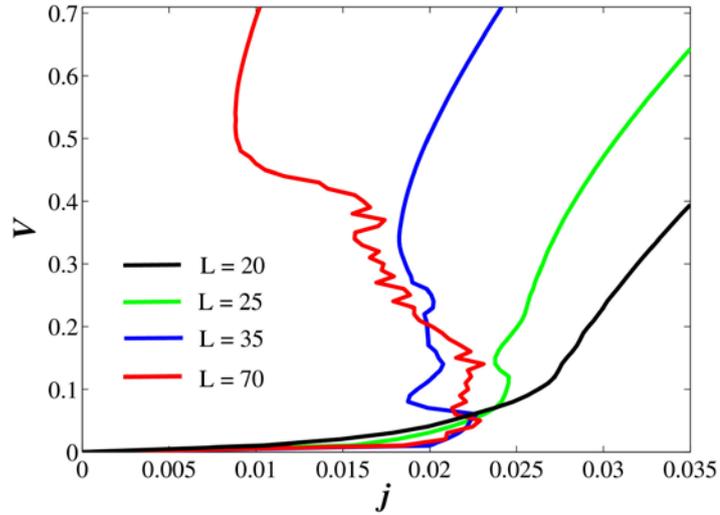

Fig. 1 (color online). The CVC of quasi-one-dimensional superconducting channels of different lengths in the voltage-driven regime.

In order to shed light on the cause, we studied the evolution of the order parameter in a channel whose length is smaller, larger, and significantly larger than the critical value of $L^{(S)}$. By evolution we mean the process showing the change of the order parameter modulus along the entire length of the specified channel with time.

We found that for the systems with $L < L^{(S)}$ for all values of applied voltage only one PSC is realized, periodically emerging in their center (fig. 2), the frequency of appearance and the size of which increase with increasing voltage.

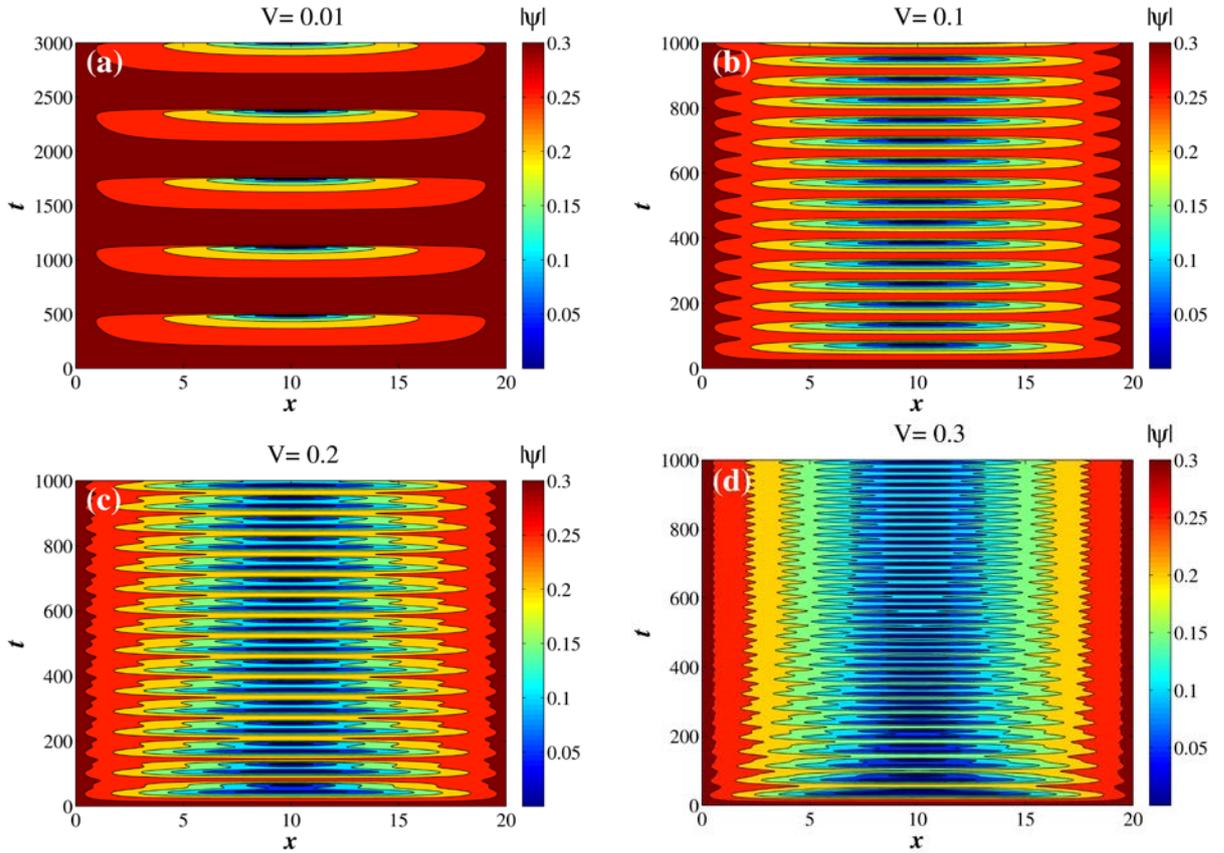

Fig. 2 (color online). The evolution of order parameter modulus in the channel with $L = 20$. The scale of the change of the order parameter modulus is shown on the right. The dark areas correspond to PSC.

For a channel with $L > L^{(S)}$ the evolution of the order parameter is a bit more complicated. To illustrate this, the channel was investigated with $L = 25 > L^{(S)}$, i.e., where CVC begins to curve and becomes S-shaped, as shown in fig. 1. As long as the voltage applied to the channel is small, there is only one PSC arising in the center of the filament (fig. 3(a)).

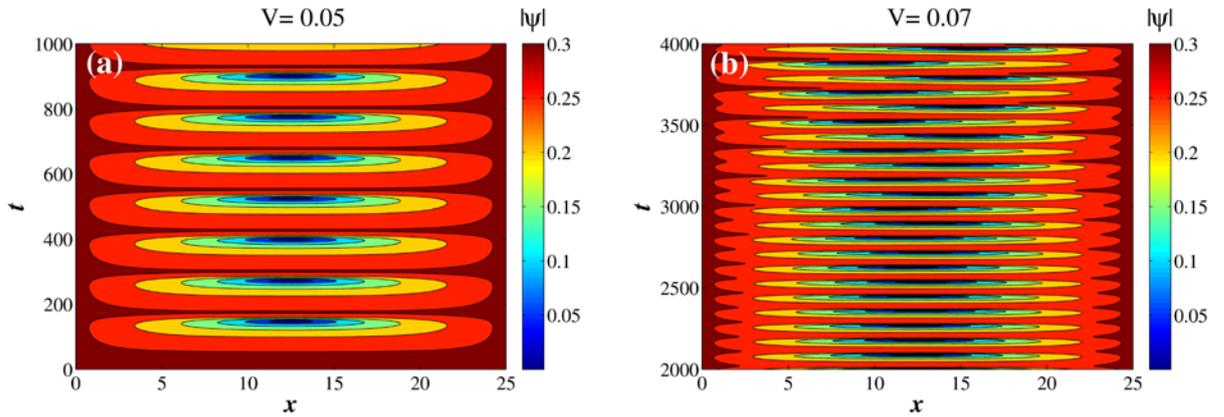

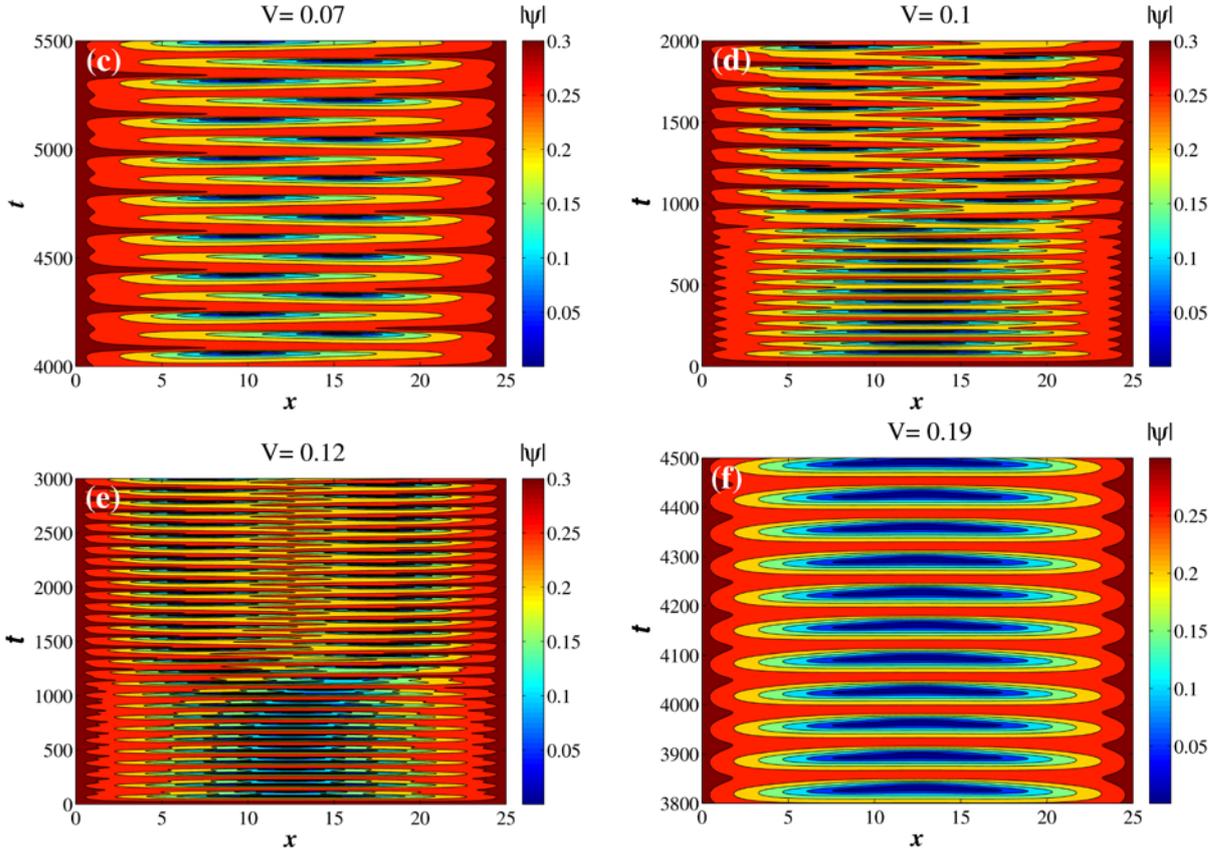

Fig. 3 (color online). The evolution of order parameter modulus in the channel with $L = 25$. The scale of the change of the order parameter modulus is shown on the right. The dark areas correspond to PSC.

However, starting from the voltage $V \approx 0.07$ and time $T_{2PSC} \approx 2900$ (see fig. 3(b)) there is a tendency for two PSC to rise in the system. Initially, the only existing PSC begins to "wobble" along the length of the channel, forming two PSC symmetrically oscillating out of phase, the distance between which increases with time and then fixates at a certain value (fig. 3(c)). Increasing the voltage to $V = 0.1$ leads to the repetition of the same scenario, i.e., "rocking" of one of the PSC with the subsequent formation of two PSC oscillating out of phase in symmetrically arranged centers of the channel halves, but now at a much earlier point in time $T_{2PSC} \approx 800$ (fig. 3(d)). By increasing the voltage to $V = 0.12$, we observed the increase of $T_{2PSC} \approx 1100$ (fig. 3(d)). The results of numerical simulations show that a further increase in the voltage increases the value of $T_{2PSC}$. Such a relationship exists until the voltage $V \approx 0.18$, after which the system returns to the existence of a single, central PSC (fig. 3(e)), increasing in size and turning into a normal domain with increasing voltage. Interestingly, for the channels with an S-shaped CVC at the initial time in the appropriate voltage range two PSC are not formed initially, they are the result of a kind of "splitting" of the central PSC.

After analyzing quasi-one-dimensional systems with $L > L^{(S)}$, we were able to make an intermediate conclusion: the beginning and end of the curvature (the exit to linear dependence) of the CVC, i.e., the upper and lower S-shaped regions, correspond to the moment of appearance and disappearance in the system of more than one PSC. We believe that this is the main reason of the S-shaped CVC.

Once the length exceeds another characteristic value $L^{(S+Oscil)} \approx 28$, in the S-shaped section of the CVC of the channel disordered oscillations begin to superimpose onto the current density in the form of sharp bends (fig. 1). With increasing length the number and the amplitude of these "deformations" increase, gradually "filling" the entire S-shaped region of the CVC. To clarify the reasons for this phenomenon, we turned again to the analysis of the evolution of the order parameter modulus in the channels for filaments whose length exceeds $L^{(S+Oscil)}$.

To start, we examined the system with $L = 35$, the length of which exceeds $L^{(S+Oscil)}$. As long as the voltage applied to the system does not exceed $V \leq 0.03$, a single PSC is present in the channel, intermittently emerging in its center (fig. 4(a)). In a range of voltages $0.03 < V < 0.07$ some time $T_{2PSC}$ later, the central PSC begins to swing, leading to the appearance of two PSC oscillating out of phase and symmetrical relative to the center (fig. 4(b)).

After the voltage reaches $V \geq 0.07$, two symmetrical PSC, located in the centers of the channel halves and oscillating synchronously, initially form in the channel (fig. 4(c)). With further increase in voltage a tendency to suppress the order parameter modulus in the central part of the system and the emergence of the third PSC are observed in the channel. Indeed, once the voltage exceeds $V > 0.17$, the third PSC forms in the center of the channel, whose dimensions vary in a periodic manner (fig. 4(d)).

When increasing the voltage to $V \approx 0.2$ (fig. 4(d)) it was found that the "side" PSC (located in the halve centers of the channel) begin to make periodic motions towards the central PSC, thus forming a central region, where the order parameter is strongly suppressed.

At $V \approx 0.23$ the offsets of these two "side" PSC become so great that at some point in time they begin to merge with the central PSC, forming a normal domain (fig. 4(e)), which increases in size with increasing voltage.

Above $V \approx 0.36$ this normal domain finally combines with the two "side" PSC, filling almost the entire channel. This moment, as in the case of the channels with length $L < L^{(S)}$, is corresponded for the exit of the CVC to linear dependence.

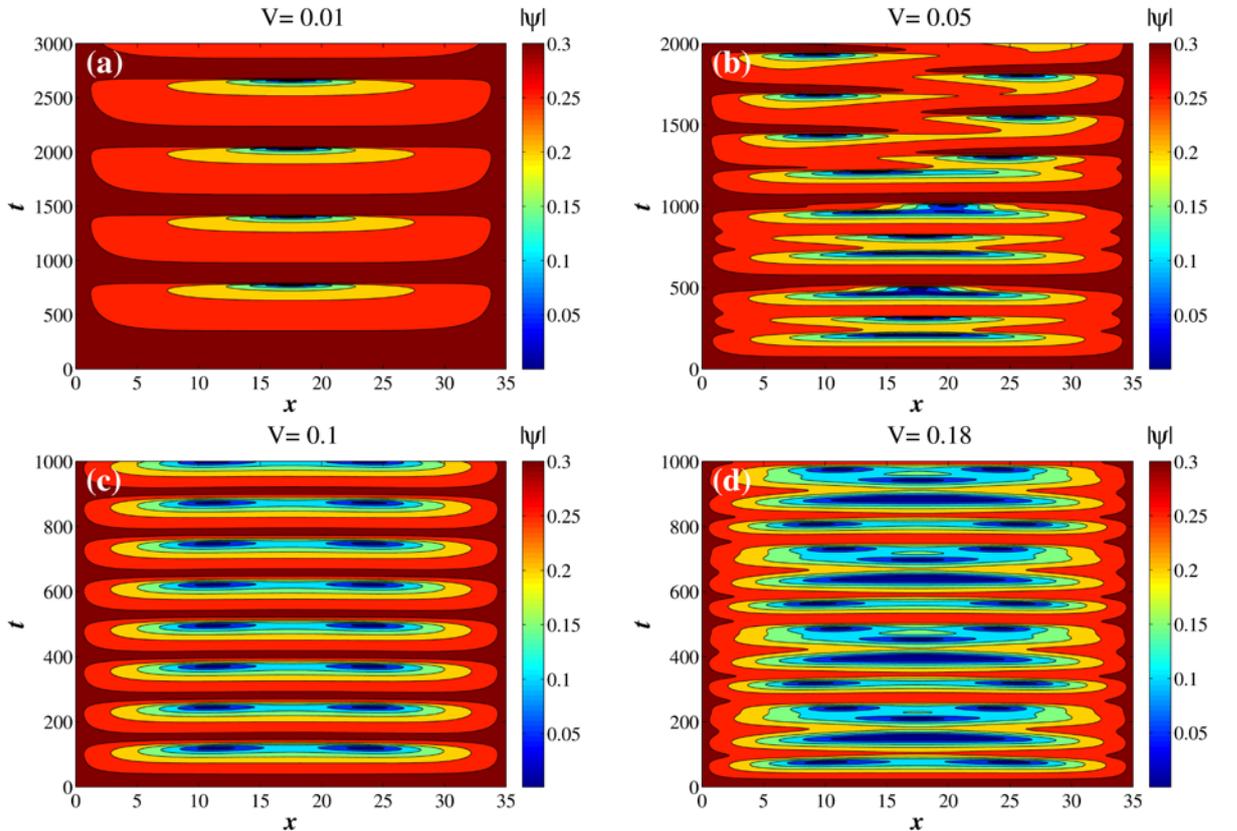

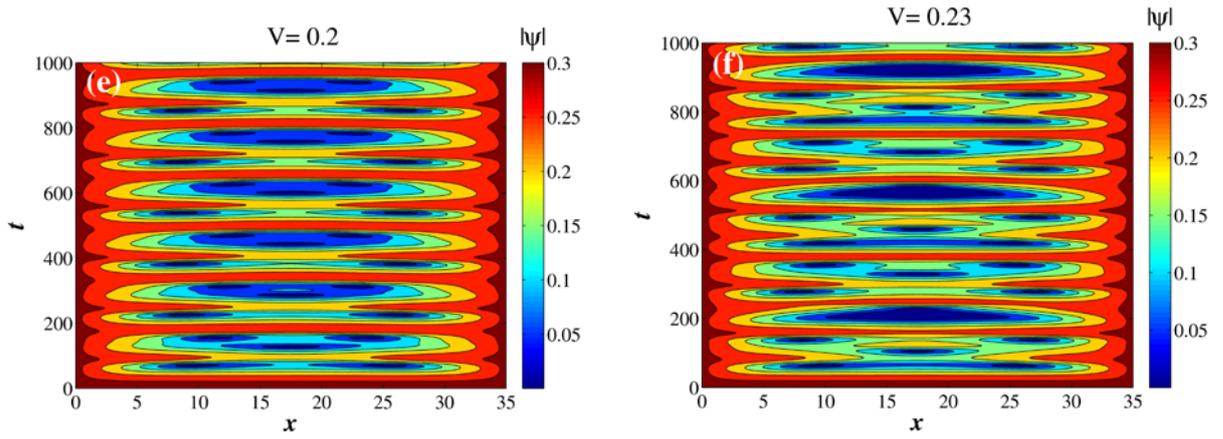

Fig. 4 (color online). The evolution of order parameter modulus in the channel with $L = 35$. The scale of the change of the order parameter modulus is shown on the right. The dark areas correspond to PSC.

It is important to note that the beginning of the curvature of the CVC of the channel with $L = 35$ corresponds to the transition from the regime with a single PSC to two symmetrically arranged PSC oscillating out of phase, which confirms the above hypothesis that the distortion of the CVC is caused by the appearance of the second PSC. In addition, note that each kink in the CVC corresponds to the voltage at which changes occur in the dynamics of behavior of the PSC, in particular, the transition to in-phase oscillations or "rocking" of the "side" PSC.

As shown by numerical simulation, the above behavior of the order parameter for $L = 35$ is qualitatively inherent to longer channels. However, if the size of a long channel is over $L > L^{(chaos)} \approx 46$, then in a certain voltage range, corresponding to the middle of the S-shaped region of the CVC, more than three PSC form, and the more the length of the channel exceeds the critical value of $L^{(chaos)}$, the more oscillating modes the new PSC have.

To illustrate the "wealth" of behavior of the order parameter in these channels we show its evolution for a system with $L = 70$.

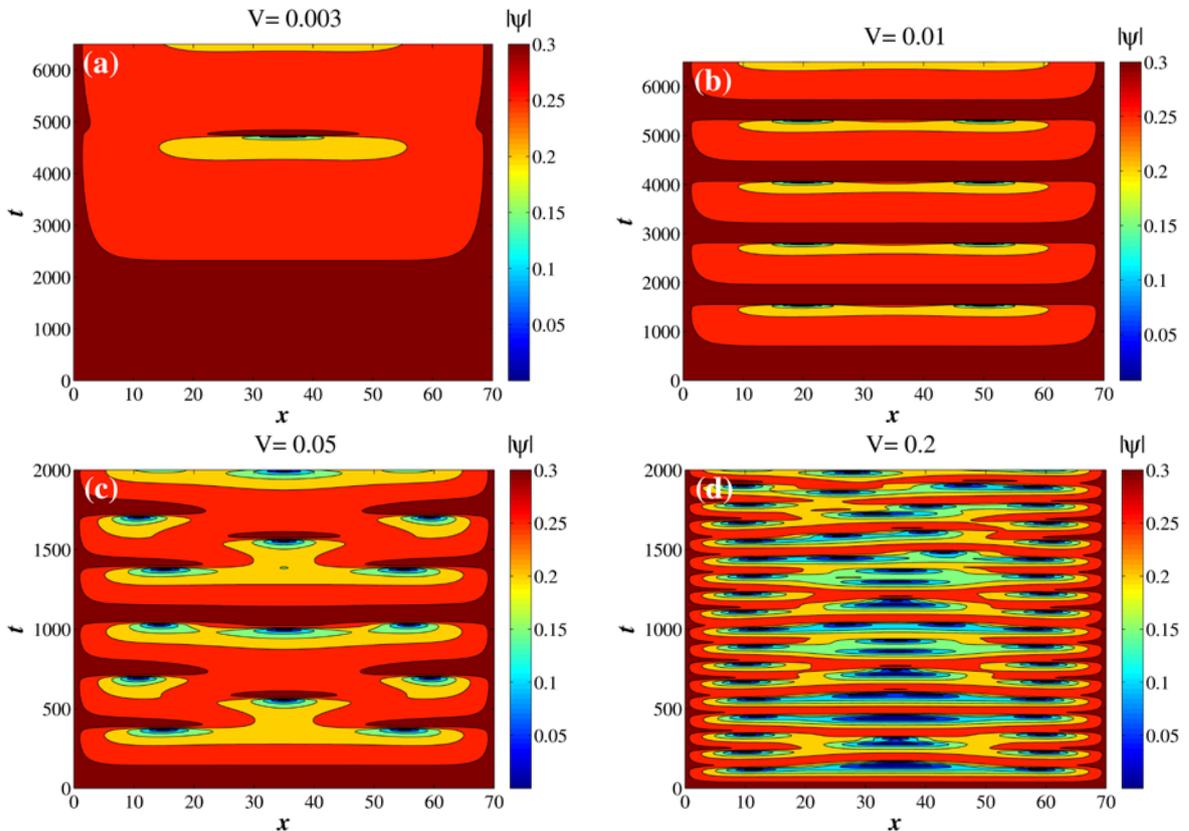

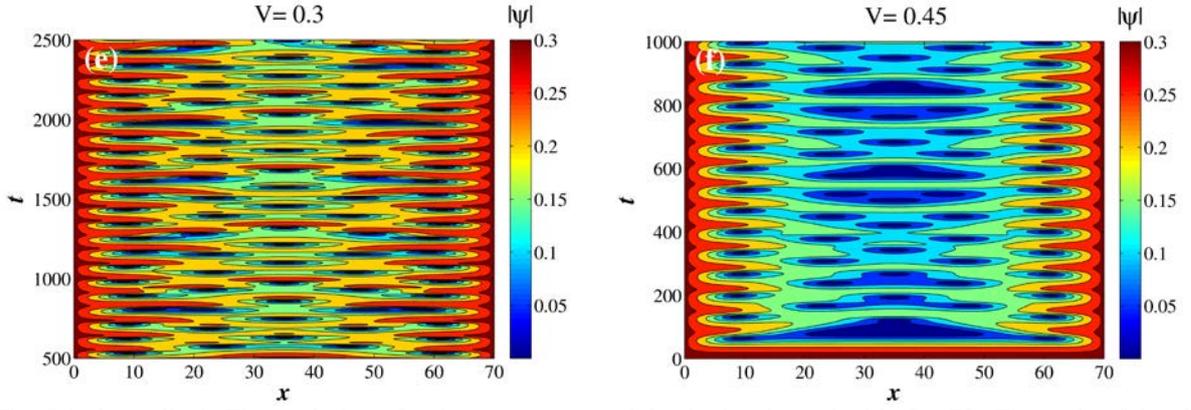

Fig. 5 (color online). The evolution of order parameter modulus in the channel with $L = 70$. The scale of the change of the order parameter modulus is shown on the right. The dark areas correspond to PSC.

According to the results of a numerical study, the formation of the first central PSC occurs in the region of very small voltages $V < 0.004$, and the smaller the voltage, the more time is required for the emergence of a single PSC (fig. 5(a)). In a narrow voltage range $0.004 \leq V < 0.035$ two PSC located at the halve centers of the channel already oscillate in the system (at first out of phase, and then in-phase) with a frequency that increases with the voltage (fig. 5(b)). The formation of the third, central PSC occurs starting from voltage $V \approx 0.035$ (fig. 5(c)), and with increasing voltage there is an increase in the oscillation frequency of the PSC and their synchronization. The appearance of new (i.e., more than three) PSC occurs when the voltage applied to the system exceeds $V \approx 0.155$. Further increase in voltage is accompanied by an increase in the number of their oscillation modes (Figs. 5(d) and 5(e)).

Such a pattern is observed up to $V \approx 0.3 \div 0.35$, after which, as the value of the voltage approaches $V \approx 0.52$, a certain order is established in the oscillations of new PSC. Now, their oscillations are almost periodic (Fig. 5(e)). Above $V > 0.52$ the system returns to the regime of three PSC, and the size of the central PSC compared to the same in the voltage range $0.035 \leq V < 0.155$ is substantially greater. Once the voltage is even greater $V > 0.72$, the central PSC, which continues to grow, absorbs the "side" PSC, forming a normal domain, which fills almost the entire channel.

It will be proven below that in a certain range of voltages, if the length of channels exceeds $L > L^{(chaos)}$, the system exhibits chaotic behavior of the order parameter. That is why to mark the transition value of the length, after which the superconducting system becomes chaotic, the superscript «chaos» was selected.

Indeed, we will now take a look at how the power spectrum of the time dependence of the current density, i.e., the parameter, which is expressed through the space-time variation of the order parameter, evolves for the channel with $L = 70$ at different voltages (fig. 6).

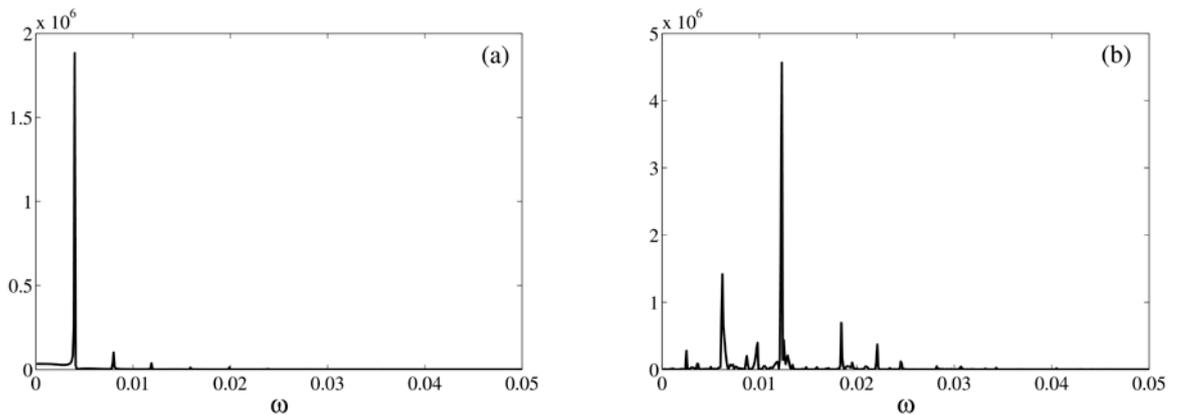

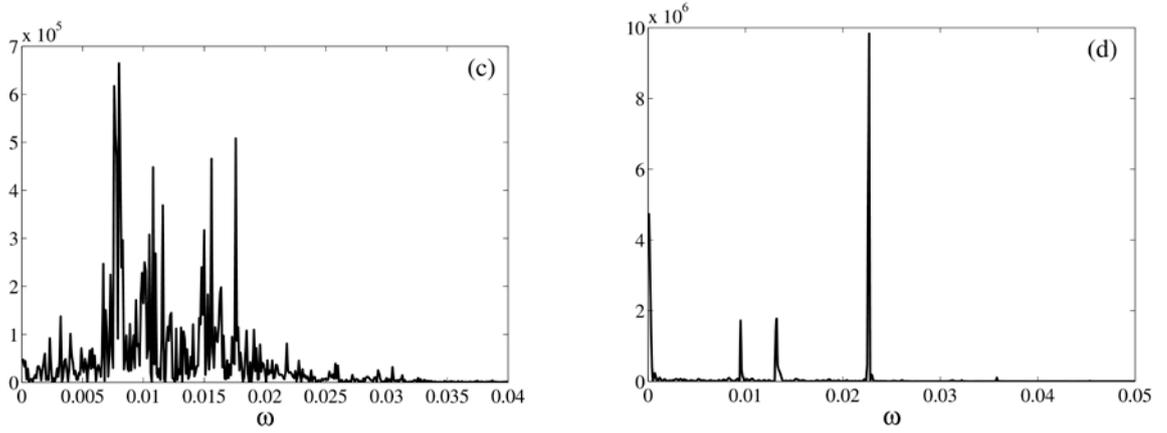

Fig. 6. Evolution of the power spectrum of the function $j(t)$ for a channel with $L = 70$ for voltages $V$ : 0.05 (a), 0.2 (b), 0.3 (c), and 0.45 (d).

Fig. 6 shows that for voltages in the range $0.155 < V < 0.52$, the power spectrum of the function $j(t)$ has no distinct single lines, which, in principle, suggests the existence of chaotic dynamics in the time dependence of the current density and the order parameter in particular.

Of course, the demonstration of the multi-peak structure of the power spectrum of current density in the voltage range corresponding to the formation of more than three PSC is insufficient to prove the chaotic behavior of the order parameter. Usually, as a criterion for the existence of chaos for a time series, which here is a set of values of the function $j(t)$, the positive-defined of the largest Lyapunov exponent is used, for the assessment of which it is necessary to define a local coefficient of "divergence" of neighboring trajectories, and then to average. The complexity of the procedure is in the determination of such a section of trajectories, where they could still be considered "close", as this section in particular is used to determine the local coefficient of "divergence". The absence of a reliable criterion of "closeness" often leads to inaccuracies in the calculation of the largest Lyapunov exponent and makes this method of detecting chaos less reliable.

Therefore, to verify the presence of chaotic behavior in a given system other tests are often employed. For example, for a quasi-one-dimensional superconducting filament in the current-driven regime, the presence of chaos was determined by means of the Poincare map [10]. In this paper, for the detection of chaotic behavior we decided to use a more reliable method — the method of calculating the correlation entropy $K_2$.

Because this feature sets the lower limit of the Kolmogorov-Sinai entropy $K$ [11] which, in turn, is equal to the average sum of the positive Lyapunov exponents (for the one-dimensional reflection, it is the Lyapunov exponent) [12] then positive-defined correlation entropy $K_2 > 0$ is a sufficient condition for the existence of chaos.

Note that direct determination of the Kolmogorov-Sinai entropy is also associated with considerable difficulties, so it is convenient to use the estimate of the Kolmogorov-Sinai entropy shown below through correlation entropy, in particular, the definition of which is easier, although it is also associated with a number of technical problems caused by the necessity of numerical calculation of limits

$$K_2 = \lim_{\varepsilon \to 0} \lim_{m \to \infty} \ln \frac{C(m,\varepsilon)}{C(m+1,\varepsilon)} \leq K, \qquad (6)$$

where $C(m,\varepsilon) = \lim_{n \to \infty} \frac{1}{n^2} \sum_{p,q=1}^{n} \theta\left(\varepsilon - \sqrt{\frac{1}{m}\sum_{s=0}^{m-1}\left[j_{p+s} - j_{q+s}\right]^2}\right)$ is the generalized correlation integral, $\theta(z)$ is the Heaviside function, $\varepsilon$ is the element of phase space, $t_p$ is the sampling time, the choice of which is important, since it determines how many values need to be considered in the summation to calculate the correlation integral with reasonable accuracy. Usually, the sampling interval is selected in such a way that each subsequent value $j_p = j(t_p)$ adds the most information or correlates with the previous one as little as possible.

We investigated the time series $j(t)$ for a wide range of lengths of superconducting channels and for voltage ranges from zero to where the CVC becomes linear. We found that in the range of voltages, for which more than three PSC arise in the system, the correlation entropy is greater than zero. This means that the specified voltage range can be clearly interpreted as an area where the dependence of current density on time, and hence the order parameter behave chaotically.

Based on the obtained data we have represented the most complete and detailed phase diagram of the resistive state of the superconducting channel in the voltage-driven regime (fig. 7).

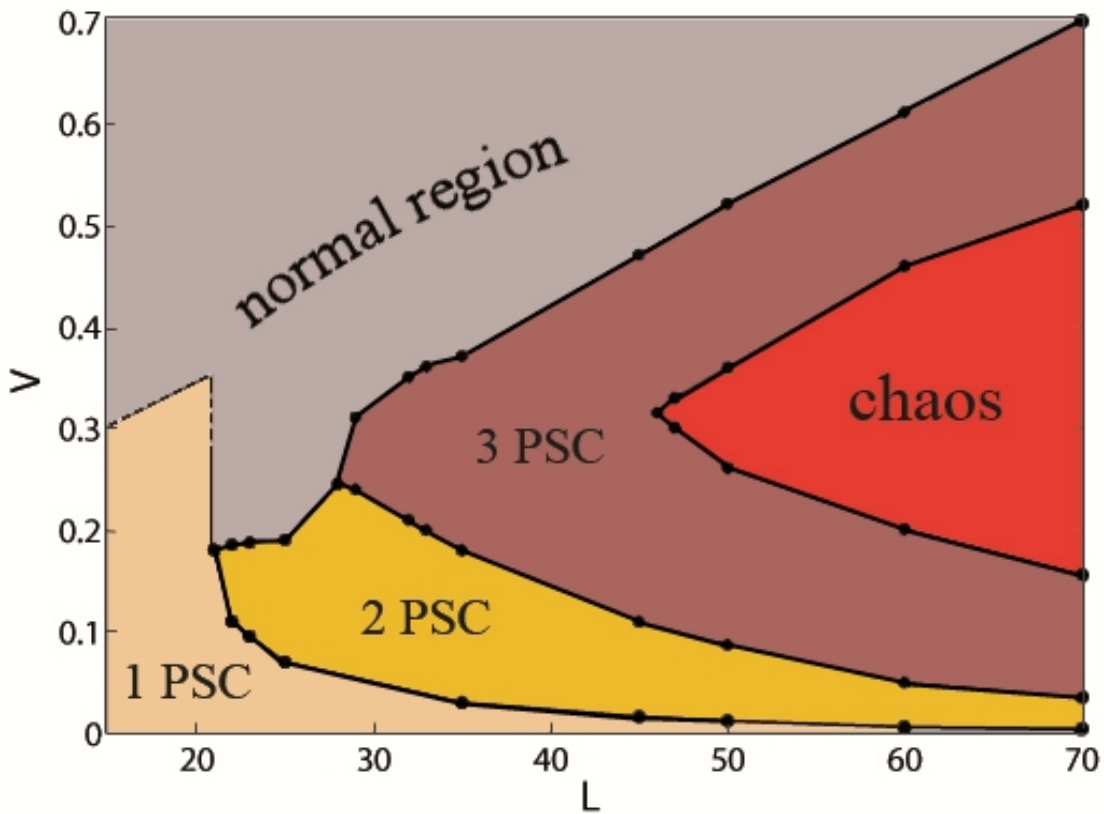

Fig. 7 (color online). Phase diagram of the resistive state of a superconducting channel in the voltage-driven regime. Each region corresponds to one, two, three PSC, or to the regime with the increasing in size normal domain. The red region defines approximately limits the state with more than three PSC. In this region, the dynamics of the order parameter are chaotic. The dashed line indicates a conditional transition for short channels from the single PSC regime to the state in which the normal domain is realized. The diagram is based on data (indicated by dots), obtained by numerical solution of the GL equations.

## Conclusions

The dynamics of the order parameter for superconducting channels of different lengths have been visualized in this paper. The evolution of this dynamic with increasing voltage was investigated, and the critical values of the lengths were determined, at which the transition occurs from the regime of one PSC in the system to the regime of two, three PSC, and so on. The CVC for a wide range of channel lengths was calculated, and it was then found that the occurrence of the S-shaped CVC and the imposed upon it "deformations" for channels with a length greater than $L^{(S+Oscil)}$ are associated with the emergence of more than one PSC and their subsequent evolutionary changes under the influence of the applied voltage.

The correlation entropy of the time dependence of the current density for channels of different lengths was calculated, and it is shown that in a certain intervals of voltages in systems that are longer than $L^{(chaos)}$, the order parameter behaves chaotically over time, leading to non-periodic behavior of the set of PSC in the central channel.

The collected data allowed us to construct the most complete phase diagram of the resistive state of a superconducting channel in the voltage-driven regime.

We note that the transport properties of quasi-one-dimensional superconducting channels have recently again attracted the attention of researchers. This is due to the fact that the coherent quantum PSC, which is the subject of theoretical studies [13] has been experimentally demonstrated in Ref. 14. Of course, the description of the quantum properties of PSC should be carried out within the framework of the microscopic theory [13] but, for a preliminary estimation of parameters of the samples, simplified phenomenological models can be useful.

As for experimental studies, to complete the picture, along with the study of the CVC it would be interesting to investigate the structures in which this quasi-one-dimensional channel is shorted by a superconducting ring. To study such PSC-based interferometers, relatively simple radio frequency technology can be used, described in detail in Ref. 15.

Indeed, in the study of flux Josephson qubits, carried out using this technique, the transition between the classical (localized states [16]) and the quantum (quantum superposition of states [17]) modes has been detected experimentally [18]. We believe that similar results can be obtained for the PSC-based interferometers.

We thank the Ministry of Education and Science in Germany (BMBF) for partial support of this work, the project UKR 10/001. Y.Y. is grateful for the support of the NAS of Ukraine through Grant No. 16-2011 for young scientists. Numerical calculations were performed on a supercomputer of ILTPE NAS of Ukraine (ds0.ilt.kharkov.ua).